\documentclass[twocolumn,showpacs,preprintnumbers,amsmath,amssymb]{revtex4}
\usepackage{tabularx}
\usepackage{amsmath,amssymb}
\usepackage{graphicx}
\usepackage{dcolumn}
\def\apj #1 #2 #3 {#1, ApJ, {\bf #2}, #3}
\def\apjl #1 #2 #3 {#1, ApJ, {\bf #2}, L#3}
\def\apjs #1 #2 #3 {#1, ApJS, {\bf #2}, #3}
\def\aap  #1 #2 #3 {#1, A\&A, {\bf #2}, #3}
\def\mnras #1 #2 #3 {#1, MNRAS, {\bf #2}, #3}
\def\pra #1 #2 #3 {#1, Phys.~Rev.~A., {\bf #2}, #3}
\def\prb #1 #2 #3 {#1, Phys.~Rev.~B., {\bf #2}, #3}
\def\prc #1 #2 #3 {#1, Phys.~Rev.~C., {\bf #2}, #3}
\def\prd #1 #2 #3 {#1, Phys.~Rev.~D., {\bf #2}, #3}
\def\pre #1 #2 #3 {#1, Phys.~Rev.~E., {\bf #2}, #3}
\def\prl #1 #2 #3 {#1, Phys.~Rev.~Lett., {\bf #2}, #3}
\def\plb #1 #2 #3 {#1, Phys.~Lett.~B., {\bf #2}, #3}
\def\science #1 #2 #3 {#1, Science., {\bf #2}, #3}
\def\nature #1 #2 #3 {#1, Nature., {\bf #2}, #3}
\def\nphysa #1 #2 #3 {#1, Nucl.~Phys.~A., {\bf #2}, #3}
\def\nphysb #1 #2 #3 {#1, Nucl.~Phys.~B., {\bf #2}, #3}
\def\nphysbs #1 #2 #3 {#1, Nucl.~Phys.~B.~Suppl., {\bf #2}, #3}

\def\h#1{\hbox{${}^{#1}$H}}

\def\h502{\hbox{$ h^{2}_{50}$}}

\def\fun#1#2{\lower3.6pt\vbox{\baselineskip0pt\lineskip.9pt
  \ialign{$\mathsurround=0pt#1\hfil##\hfil$\crcr#2\crcr\sim\crcr}}}
\begin{document}
%
\title{Observational Constraints on Accelerating Brane Cosmology with Exchange between the Bulk and Brane}
\author{K. Umezu$^{1}$, K. Ichiki$^{1,2}$, T. Kajino$^{1,2}$}
\author{G. J. Mathews$^{1,4}$}
\author{R. Nakamura$^{3}$}
\author{M. Yahiro$^{1,3}$}

\affiliation{
$^1$National Astronomical Observatory of Japan, and Graduate University for Advanced Studies, 2-21-1 Osawa, Mitaka, Tokyo 181-8588, Japan}
\affiliation{
$^2$Department of Astronomy, Graduate School of Science, University of Tokyo, 7-3-1 Hongo, Bunkyo-ku, Tokyo 113-0033, Japan }
\affiliation{
$^{3}$Department of Physics, Graduate School of Science, Kyushu University, 6-10-1 Hakozaki, Higashi-ku, Fukuoka 812-8581, Japan }
\affiliation{
$^{4}$Center for Astrophysics, Department of Physics, University of Notre Dame, Notre Dame, IN 46556, U.S.A. }
\date{\today}
\begin{abstract}
 We explore observational constraints on a cosmological brane-world scenario in which 
the bulk is not empty.  Rather, exchange
of mass-energy between the bulk  and the bane is allowed. 
The evolution of matter fields to an observer on the brane is 
then  modified due to new  terms in the energy momentum
 tensor describing this exchange.
We investigate the constraints
from various cosmological observations on the flow
of matter from the bulk into the brane.
Interestingly, we show that it is possible to have
a $\Lambda = 0$ cosmology to an observer in the brane which satisfies standard cosmological constraints including the CMB temperature fluctuations, Type Ia supernovae at high redshift, and the matter power spectrum.
This model even  accounts for
the observed suppression of the CMB power spectrum at low multipoles.
In this cosmology, the observed cosmic acceleration is attributable
to the flow of matter from the bulk to the brane.
A peculiar aspect of this cosmology is that the 
present dark-matter content of the universe
may be significantly larger than that of a
 $\Lambda$CDM cosmology.  Its influence, however, is
offset by the dark-radiation term.  
Possible additional observational tests of this new cosmological 
paradigm are suggested.
 \end{abstract}
\maketitle
%
%
%

\section{INTRODUCTION}

A puzzling question in modern cosmology has been 
the nature and  origin 
of the dark energy responsible for the apparent acceleration 
\cite{garnavich}
of the universe in the present epoch.
The simplest explanation is, perhaps, that of a cosmological constant,
or a vacuum energy in the form of a "quintessence" scalar field
slowly evolving along an effective potential.
The existence of a cosmological constant or quintessence,
 however,  leads inevitably to the well known
  fine tuning problem as to
why the  present dark  energy is so small compared to
the natural scales of high-energy physics.
There is also a cosmic coincidence problem
as to why the universe has conspired
to produce nearly equivalent energy contents in matter and dark   energy
at the present time.  Moreover, most quintessence models
are now ruled out \cite{WMAP} by the fact that the equation of state parameter,
$w \equiv p/\rho$ is so close to -1.  This
 implies that there is little evidence for
evolution in a quintessence field  so that a cosmological constant remains the most likely interpretation.  

Accounting for such dark energy, however, poses an  
unsolved theoretical challenge.
Ultimately,  one would hope that the existence
of this dark energy could be
accounted for in the context of string theory or some other
unified theory.
This problem is
exacerbated, however, by the fact
that it appears difficult 
to accommodate a cosmological constant in string theory
\cite{hellerman}, although some progress has been made 
\cite{banks}.  Recently, it has been proposed \cite{kolb1,kolb2} that the observed
cosmic acceleration may be  an artifact of inhomogeneities in the distribution of mass-energy.  That possibility, though compelling
for its simplicity, has not yet been established \cite{Chung,Hirata}.

  Hence, in this paper, we consider it worthwhile to analyze 
  an alternative  mechanism by which
the observed cosmic acceleration could be produced even
without the need to invoke dark energy and its associated
complexities.
Specifically, we explore models  in which
the cosmic acceleration is driven \cite{Kiritsis,Tetradis,TetradisI,Myung,TetradisII, Tetradis05a}
 by the flow of dark matter from
a higher dimension  (the bulk)
into our three-space (the brane).

This study is also in part motivated by the currently  popular view
that our universe could be a sub-manifold embedded
in a higher-dimensional space-time. 
This paradigm derives from the  low-energy
limit of the heterotic M-theory \cite{Horava} which becomes
an 11-dimensional supergravity compactified on a line segment 
to two 10-dimensional $E8\times E8$ gauge theories.  
Hence, the universe can be envisioned as
 two smooth 10-dimensional manifolds 
(9-branes) embedded in a bulk dimension. 

A thin three-brane (Randall-Sundrum brane) 
embedded in a five-dimensional anti-de Sitter space $AdS_5$ 
has been proposed as a practical phenomenological model \cite{Randall} in which to explore such higher dimensional physics. 
This approach poses an alternative to the standard Kaluza-Klein (KK)
compactification of extra dimensions, through the localization of 
the graviton zero mode on the brane.
This brane approach also provides a new way of understanding 
the hierarchy between the four-dimensional Planck scale $M_{\rm pl}$ 
and the electro-weak scale \cite{ArHa,KaSc}.

Of relevance to the present study is that in this
phenomenological approach, one can relax the requirement that
the gauge groups be rigorously confined to the brane.
For example, particles might  be localized on a defect 
in the higher dimensional space \cite{Rub83,Akama:1982jy}.
The simplest example of this would be a domain wall (brane) 
in (4+1) dimensions.
In such a picture, the extra (bulk) dimension can be infinite
and the observed 3-dimensional fields are 
represented by zero modes of 
bulk fields in the domain-wall background.  These modes are localized
and thus behave like 3-dimensional mass-less fields.  
In this domain-wall scenario, 
physical matter fields are dynamically confined to this sub-manifold,
while gravity can reside in the extra bulk dimension.

The stability of massive matter fields has been 
analyzed \cite{dubovsky} in this 
scenario where it was  shown that such massive particles 
are metastable on the brane.  That is, for both scalar and fermion
fields, the quasi-normal modes are metastable states that decay into
continuum states.  From the viewpoint of an observer in the
4-dimensional space time, these massive particles appear to propagate
for some time in three spatial dimensions and then disappear into the
fifth dimension.
This disappearance  of massive particles from the 3-brane
constitutes  an energy flow from the brane to the 
bulk. The cosmological constraints on such disappearing
 matter have been studied in Ref.~\cite{DDM}.

Moreover, if the massive particles can also exist in the bulk, 
it becomes  possible to 
consider the inverse flow from the bulk into our three-brane.
In this paper, we build a model with such  mass-energy exchange, 
in which 
the flow from the bulk to the brane provides the present observed
acceleration \cite{Kiritsis,Tetradis,Myung,TetradisI,TetradisII, Tetradis05a} of the universe.   As noted in Ref.~\cite{DDM} a heavy ($\sim$TeV) dark-matter candidate
such as the lightest supersymmetric partner is likely to have
the largest tunneling rate between the brane and bulk.
Moreover, a recent study \cite{Tetradis05b} of energy transfer from the
bulk through parametric resonance indicates that massive
particles are most efficient for depositing energy on the brane.
Hence, in the present work we mainly consider the exchange
between the bulk and brane involving a growth of the  cold
dark-matter component on the brane.
 This model 
[which we refer to as growing cold dark matter (GCDM)]
is, thus, an alternative 
to the standard $\Lambda$ plus cold dark matter (S$\Lambda$CDM) cosmology to an observer on the
3-brane. 

 We test this model by analyzing 
the observations of Type Ia supernovae at high redshift,
the temperature fluctuation spectrum of the cosmic microwave background (CMB), and the matter
power spectrum.  Surprisingly, all of these constraints 
can be satisfied in this model even without introducing 
a cosmological constant on the brane.
A shocking feature of the best fit model, however,
 is that
the present dark-matter content becomes 
as much as an order of magnitude
larger than that of a  $S\Lambda$CDM cosmology.
The influence of this excess dark matter, however,  is suppressed  by
the large dark-radiation component associated with the flow
of matter from the bulk dimension.  
In fact, this model provides a natural explanation for
the observed suppression of the CMB power spectrum for
the lowest multipoles due to the effect of the late arriving
dark matter on the integrated Sachs-Wolf effect.  We show
here that the matter power spectrum exhibits
the correct  power on large  and small scales, however, a somewhat 
large bias factor is required. 

\section{Model with Energy Exchange between the Bulk and Brane }

We begin with the five-dimensional Einstein equations
\begin{equation}
G_{AB}= \kappa^2 T_{AB}, 
\end{equation}
where $\kappa^2=M^{-3}$  is the 5-dimensional gravitation
constant with $M$ the 5-dimensional Planck mass.

$T_{AB}$  is the five-dimensional total energy-momentum tensor.
We use the index notation $A,B=(0,1,2,3,5)$
to clearly distinguish the bulk fifth dimension. 
The metric is assumed to be of the Randall-Sundrum (RS) \cite{Randall} form.
For simplicity, however, we 
consider a coordinate system with a static bulk dimension and  an expanding brane or moving \cite{Krauss} brane.  
Hence, we choose a  metric in Gauss-normal coordinates 
\begin{equation}
ds^{2}=-\hat n^{2}(t,y) dt^{2}+\hat{a}^{2}(t,y)\gamma_{ij}dx^{i}dx^{j}
+ dy^{2}~~,
\end{equation}
where $\gamma_{ij}$ is a maximally symmetric 3-metric. 
As an illustrative model, we allow the Hubble expansion of the brane relative to a static
bulk to produce a nonzero 5-component of the  bulk fluid velocity as discussed below.

\subsection{Energy-Momentum Tensor}
When matter is present in the bulk, the evolution of the
brane is not autonomous.  It becomes  necessary to explicitly
define the bulk energy-momentum tensor.
The general bulk energy-momentum tensor for a perfect fluid 
in this 5-dimensional space-time is,
\begin{equation}
T^A_{~B} = (\tilde \rho + \tilde p) U^A U_B + \delta^A_{~B} \tilde p ~~,
\end{equation}
where $\tilde \rho$ and $\tilde p$ denote density and pressure
in 4-dimensional space.
In Gaussian-normal coordinates, however, 
the bulk 5th dimension is fixed so that the four density and pressure
are proportional to the three-volume like ordinary
three-space quantities.
 For normal matter on the brane we denote the density and pressure
as $\rho$ and $p$.  In our model in which dark matter can exist both on the
brane and in the bulk we denote the dark-matter  three density and pressure on
the brane as $\bar \rho$ and $\bar p$. Since the density, pressure and equation
of state can be different in the bulk we denote the four density of dark matter
in the bulk as $\hat \rho$ and $\hat p$.  

In a frame for which the three-brane is at rest we can decompose
the energy-momentum tensor into three parts.  
\begin{equation}
T_{AB} = \delta(y)\biggl(T_{AB}^{BRANE}\biggr) + T_{AB}^{BULK} + T_{AB}^{DM}~~,
\end{equation}
where the $\delta$-function identifies the location of the
 brane at $y=0$ to which the standard-model particles are 
 assumed to be  confined.
 $^{(BRANE)}T^A_{~B}$, corresponds to 
the  usual three-density and pressure, $\rho$ and $p$, of ordinary 
relativistic and non-relativistic particles plus
the tension $\tau$ on the brane.
\begin{equation}
 ^{(BRANE)}T^A_{~B} =
{\rm diag}(-\tau-\rho,-\tau+p,-\tau+p,-\tau+p,0)~~.
\end{equation}
The bulk is taken to be $AdS_5$, 
so that the five-dimensional cosmological constant $\Lambda_5$ is 
negative in the bulk.  Hence,
 we write the vacuum energy-momentum
for the bulk, $^{(BULK)}T^A_{~B}$, as
\begin{equation}
^{(BULK)}T^A_{~B} =
{\rm diag}(-\Lambda_5,-\Lambda_5,-\Lambda_5,-\Lambda_5,-\Lambda_5).
\end{equation}
In our model dark matter is 
allowed to exist both on the brane and in the bulk. 
Hence, we decompose the dark-matter energy-momentum tensor, $^{(DM)}T^A_{~B}$,
 into the usual three-density $\bar{\rho}$  and pressure 
$\bar{p}$ of dark matter on the brane, plus the  
bulk components $^{(DM-BULK)}T^A_{~B}$.
 \begin{equation}
^{(DM)}T^A_{~B} =
\delta(y){\rm diag}(-\bar{\rho},\bar{p},\bar{p},\bar{p},0)
+^{(DM-BULK)}T^A_{~B}.
\end{equation}
where,
\begin{equation}
^{(DM-BULK)}T^0_{~5} \sim (\hat \rho + \hat p)U_5 ~~,
\end{equation}
represents  the 
matter-energy flow from the bulk to the brane,
while 
\begin{equation}
^{(DM-BULK)}T^5_{~5} = (\hat \rho + \hat p)U^5U_5 + \hat p ~~,
\end{equation}
represents a bulk pressure in the limit of vanishing $U_5$.

As we shall see, the cosmology  for an observer living on the 
three-brane will
depend upon the equation of state (EOS) properties for matter in the bulk dimension.  It is not guaranteed, however, that matter will
have the same EOS properties in the bulk as on the brane.  Inevitably, assumptions must be made \cite{Tetradis, vandeBruck}
about the form of the energy-momentum tensor for matter in the bulk.  For example, a bulk field \cite{Chamblin}, or
radiation emitted from the brane to the bulk \cite{Langlois}
have been considered along with possible thermal effects \cite{ChamblinT}.  In a string theory sense, particles in the bulk may appear as a topological defect between two branes.
For the purposes of this paper, we adopt \cite{Kiritsis,Tetradis}  the usual parametrization of the EOS for matter in the bulk.  Scaling
the bulk dark-matter density to the present critical density
$\rho_{cr}$ on the brane, we write:
\begin{equation}
\hat \rho \propto 
\biggl(\frac{\rho_{cr}}{ a^{q}}\biggr)~~,
\label{scale1}
\end{equation}
where $a(t)$  is the scale factor on the 
brane, defined as 
$a(t)=\hat a(t,0)$, and for a fixed bulk dimension
the scaling parameter $q$ can be written $q = 3(1+\hat w)$, where
$\hat w = \hat p/\hat \rho$ is the usual equation of state parameter with,
$\hat w = 0$ for normal cold dark matter, while
$\hat w = 1/3$ for relativistic matter.   A value of $\hat w=-1$ corresponds 
to a vacuum energy, while $\hat w = -2/3$ for a string-like topological defect.

To illustrate this cosmology we consider two models
for the motion of matter in the bulk relative to the brane.
In the Gaussian-normal coordinates, the five velocity
of matter in the static bulk with respect to the expanding 3-brane 
is just \cite{Mukohyama:1999wi}
\begin{equation}
U_5 \propto  -l H~~,
\end{equation}
where
\begin{equation}
l = [-6 M^3/\Lambda_5]^{1/2}~~,
\end{equation}
is the bulk curvature radius \cite{DDM}.
  
  As a second illustration
we also consider the case of a 
constant $U_5$ component.
 Such a model
might correspond, for
example, to  gravitational accretion of
matter from the bulk toward the brane at a constant rate.

For the purposes of  scaling, we parameterize the 0-5 component
 of the bulk dark-matter energy-momentum tensor as
\begin{equation}
^{(DM-BULK)}T^0_{~5} =  -\frac{\alpha}{2} 
\biggl(\frac{\rho_{cr}}{a^{q}}\biggr) \times H~~, 
\label{scale2}
\end{equation}
where we have assumed a constant transition rate of
matter between the bulk and the brane absorbed into the dimensionless
parameter $\alpha$ which can be either  positive or negative.
For the case of constant $U_5$ we replace $H \rightarrow H_0$
in the above scaling, where $H_0$ is the present Hubble parameter.

\subsection{Accelerating Cosmologies}
The cosmological equations of motion with brane-bulk energy exchange have previously been formulated 
in Refs. \cite{Kiritsis,Tetradis, Myung, TetradisI,TetradisII, Tetradis05a}.  The covariant derivative of the energy-momentum tensor
 leads to the usual energy conservation condition for various components $i$ 
 of normal matter on the brane;
\begin{equation}
\frac{\dot\rho_i}{\rho_i}+3(1+w_i)\,{\frac{\dot a}{a}}  = 0~~,
\label{rho}
\end{equation}
plus a new condition for the dark matter which takes into account
the flow of matter from the brane world,
\begin{equation}
\frac{\dot{\bar{\rho}}}{\bar \rho}+3(1 + \bar w){\frac{\dot a}{a}} =
-\frac{T}{\bar \rho},
\label{rho-DM}
\end{equation}
where $T \equiv 2T^0_{~5}$ is the discontinuity of the (0,5) 
component of 
$^{(DM-BULK)}T^A_{~B}$ at $y=0$. 
The (0,0) component of the Einstein equation 
 produces a modified Friedmann cosmology to an
 observer on the brane:\\
\begin{eqnarray}
H^2 = \frac{\dot a^2}{a^2}&=&\frac{8}{3}\pi G_N(\rho+ \bar \rho)-\frac{k}{a^2}+\Lambda_4 \nonumber \\
&&+\frac{\kappa^4}{36}(\rho+\bar \rho)^2+\chi  ~~,
\label{a}
\end{eqnarray}
while the (5,5) component leads to an equation for the
dark radiation term $\chi$,
\begin{eqnarray}
~
&&\dot\chi+4\,{\frac{\dot a}{a}}\,(\chi+ \frac{\kappa^2}{6} T^5_5)=\frac{\kappa^4}{18}
\left( \rho+\bar \rho +\tau \right)T~~.
\label{chi}
\end{eqnarray}
In Eqs.~(\ref{a}) and (\ref{chi}) the usual change  variables has been introduced:
\begin{equation}
^{(DM-BULK)}T^5_{~5} \equiv T^5_{~5}~~,
\end{equation}
\begin{equation}
\Lambda_4 \equiv (\kappa^2/6) 
\biggl[\Lambda_5+(\kappa^2/6)\tau^2 \biggr]~~, 
\end{equation}
and 
\begin{equation}
\kappa^4 \tau/18 \equiv 8\pi G_N/3~~.
\end{equation}
 Introducing these identities into Eq.~(\ref{a}) leads to the usual 
Friedmann equation plus  extra terms with $\chi$ and $(\rho+\bar \rho)^2$. This 
modified Friedmann equation 
can be normalized in the usual way so that the
spatial curvature takes on values $k=-1,0,1$.
In the limit of an empty bulk and no exchange between the bulk and the brane, 
 $T^0_5=T^5_5=0$, the quantity 
$\chi$ varies with scale factor on the brane as $C/a^4$. Thus, $\chi$ reduces to the standard
dark-radiation term when all matter fields
are confined to the brane. 

Now we assume that the dark matter is cold on the  brane: $\bar p=0$.
First, we consider the special case $T=\Gamma \bar \rho$. 
A solution of Eq.~(\ref{rho-DM}) 
is $\bar \rho= C \exp{(-\Gamma t)}/a^3$. 
This is precisely the disappearing dark 
matter cosmology introduced in our previous paper \cite{DDM}. 
Thus, $T=\Gamma \bar \rho$ 
describes  matter-energy flow from the brane into the bulk.
In this cosmology dark-matter particles can exist  in both the brane and the bulk. 
Hence, negative  $T$ describes a reverse flow from 
the bulk to the brane. 
However, in the present paper, we only consider the inflow from the bulk into the brane.

 \subsubsection{An Illustration}
 To see how an accelerating cosmology arises, consider
 the simple 
example of a constant flow rate from the bulk into the brane,
i.e. $T=-\Sigma$ for a positive constant $\Sigma$. 
In the case of $k=0$ and $\rho << \bar \rho$, (nearly realized 
in the present universe) equations (\ref{rho-DM})-(\ref{chi}) have 
a fixed point $(H_*,\bar\rho_*,\chi_*)$ governed by,
\begin{eqnarray}
&&3\bar{\rho_*} H_* = \Sigma,
\\
&&H_*^2=\frac{8}{3}\pi G_N \bar{\rho_*} +\Lambda_4+ \frac{\kappa^4}{36} \bar{\rho_*}^2+ \chi_*,
\label{H*}
\\
&&4 H_* (\chi_*+ \frac{\kappa}{6} T^5_5)= - \frac{\kappa^4}{9} \tau \Sigma  ~~.
\end{eqnarray}
The constant r.h.s.~of equation (\ref{H*}) then is manifestly equivalent
to a cosmological-constant dominated universe.   

In general, the fixed point remains even 
in the case of $\Lambda_4=0$. Thus,  the present accelerating 
universe need not be attributed  to $\Lambda_4$ at all,
 but could also arise from 
constant bulk components of the five dimensional
stress-energy tensor,  $\Sigma = -2T^0_5$ and $T^5_5$.
When a fluid is static, 
$T^5_5$ represents a  pressure, it is then natural to set
the pressure to be zero. But $T^5_5$ is in general 
non-vanishing for a moving flow.
 Another possibility for non-vanishing $T^5_5$ is 
that the dark matter in the bulk is in a vacuum state. In this case, $T^5_5$ is a vacuum pressure, and it is likely that $\Sigma=0$.

\subsection{$\Lambda_4 = 0$ Accelerating cosmology}
We consider two growing dark matter models of interest to the present 
work.
For the case of vanishing $\Lambda_4 $ and constant $U_5$,
 the evolution equations of energy densities are rewritten as,
\begin{eqnarray}
\label{rho1}
&&\dot{\bar{\rho}} + 3H\bar{\rho} = \alpha/a^q \times \rho_{cr} H_0,
\\
&&\dot\rho + 3H(\rho+p) = 0 ~,\\
&&\dot\rho_\chi + 4H\rho_\chi = -\alpha/a^q \times \rho_{cr} H_0,
\end{eqnarray}
where $\rho_\chi = {\chi}/(8\pi G_N/3)$ is the dark-radiation term
which plays the role of dark energy.
For the case of $U_5= -l H$,  the evolution 
equations for the energy densities are
\begin{eqnarray}
&&\dot{\bar{\rho}} + H\biggl[3\bar{\rho} - \alpha/a^q \times \rho_{cr}\biggr] = 0~,\\
&&\dot\rho + 3H(\rho+p) = 0 ~,\\
&&\dot\rho_\chi + H\biggl[4\rho_\chi + \alpha/a^q \times \rho_{cr} \biggr]
= 0~.
\label{rhochidot}
\end{eqnarray}
These equations satisfy the conservation's law on the brane.
Note, that in this latter case, the fact that $\rho_\chi < 0$
means that Eq. (\ref{rhochidot}) quickly evolves to 
 $\dot\rho_\chi = 0$ for the case when $q = 0$.
 Indeed, as long as $q < 3$ an accelerating cosmology
 eventually emerges \cite{Tetradis}.
Hence, the dark radiation contribution becomes constant and
indistinguishable from a cosmological constant.

For the $\Lambda_4 \equiv \Lambda = 0$ accelerating
cosmology of interest here, the modified  Friedmann equation 
for cosmic expansion can be written
\begin{equation}
 H^2 = \left(\frac{\dot a}{a}\right)^2 
     = \frac{8 \pi G}{3}(\bar\rho + \rho + \rho_\chi) +\frac{k}{a^2}~.
\label{Friedmann}
\end{equation}
Here we have neglected the $\rho^2$ term. 
This term decays
rapidly as $a^{-8}$ in the early radiation dominated epoch and hence
is insignificant for the present studies.
Note, that Eq. (\ref{Friedmann}) corresponds to the limit in which $T^5_5 \approx 0$,
and $\tau >> MAX(\rho, \bar\rho)$.

\begin{figure}[t]
\begin{center}
\rotatebox{-90}{\includegraphics[width=7cm]{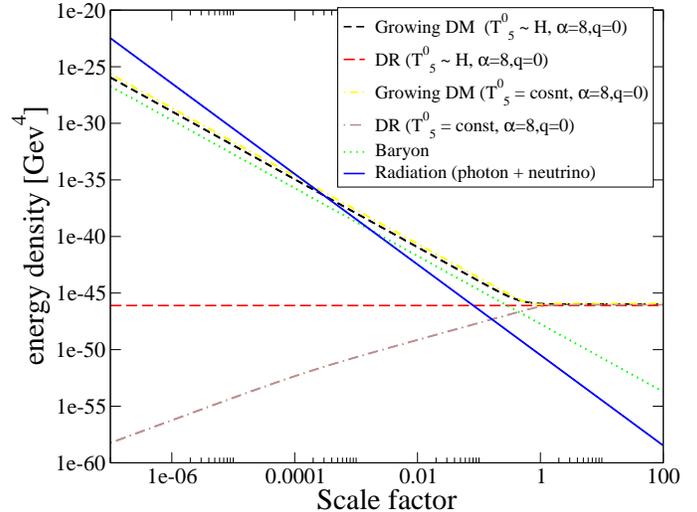}}
\end{center}
\vspace{-0.5cm}
\caption{Evolution of various
energy densities as a function of scale factor in a
$\Lambda = k = 0$
 growing cold-dark-matter model. 
 The curves labeled as "Growing DM" shows the dark radiation component.
Note that the
 dark radiation content is plotted as its absolute value.
 This quantity is actually negative in these models.
 }
\label{fig:energy}
\end{figure}

Figure \ref{fig:energy} illustrates  the evolution of 
various components on the brane in a simple 
$\Lambda=k=0$ cosmology. 
This cosmology separates into three characteristic
epochs on this figure. First, the usual early radiation dominated
epoch ($a < 10^{-4}$). Second, a dark-matter dominated epoch
($10^{-4} < a < 10^{-1}$). Third, the dark matter and dark radiation
dominated accelerating epoch ($a >0.5$).

Both of our two models for the growth rate of dark matter are illustrated on Figure \ref{fig:energy} for the case of 
$q=0$, $\alpha = 8$.  During the first and second epochs, the dark
radiation component for the case of constant $U_5$ evolves as $\rho_{DR} \propto a^{2-q}$ and $\rho_{DR}
\propto a^{3/2-q}$, respectively.  
For the
case of brane expansion into the bulk, $U_5 \propto H $,
the dark radiation
term is initially much larger than for the case of constant $U_5$
and it remains at a nearly constant value.  
Nevertheless, during the radiation- and matter-dominated epochs,
the dark radiation term for both models is
 insignificant until recent history near $z \approx 1$.  Hence, this cosmology
 is not constrained by primordial nucleosynthesis. \cite{Ichiki:2002eh}

This model thus explains the smallness
problem of the apparent cosmological constant as simply due to
the slow tunneling of matter from the bulk onto the brane.
The cosmic acceleration only occurs now, because this is the epoch
for which the fixed-point solution could be obtained.
  Moreover, in what follows we show that the fits to observations based upon these two models are indistinguishable from each other and from the best standard
$S\Lambda$CDM model.

Eventually, in the
third region the cosmology of interest to this paper emerges.
The 
dark radiation component dominates.  In the case of
$q\approx 0$ it becomes a constant dark-energy density. Hence, the dark radiation associated with
in-flowing  cold dark matter leads the cosmic
acceleration without the need for a cosmological constant 
on the right hand
side of Eq. (\ref{Friedmann}) as long as matter in the bulk has the
right equation of state. 

\section{Observational Constraints}
 Having defined the cosmology of interest we now analyze the
 various cosmological constraints as a test of this hypothesis by solving
 equations (24)-(30) numerically. In particular, we examine the
 magnitude-redshift relation for type Ia supernovae (SNIa), the cosmic microwave
 background (CMB) and the matter power spectrum $P(k)$. Parameters summarizing
 the best fits to these constraints are given in Table 1 for models with and
 without growing cold dark matter.  The S$\Lambda$CDM fits are consistent with
 the usually inferred cosmological parameters (e.g. \cite{WMAP}. An important
 constraint on the GCDM model is that the sum of $\Omega_{DM}+\Omega_{DR}$ be
 approximately equivalent to the sum of $\Omega_{DM}+\Omega_{\Lambda}$ in the
 standard cosmology.  This is evidenced by the fifth column of Table 1.   Note,
 however, that the magnitude of $\Omega_{DM}$ and $\Omega_{DR}$ individually can
 be quite large in the GCDM model, as long as $\Omega_{tot} \approx 1$.

\section{Supernova Likelihood analysis}
The apparent brightness of the Type Ia supernova standard candle with
redshift is given \cite{Carroll} by a simple relation which we slightly
modify to incorporate the brane-world cosmology given in
Eq.~(\ref{Friedmann}). The luminosity distance becomes,
\begin{eqnarray}
D_L &=& \frac{c (1+z)}{ H_0  \sqrt{\Omega_{k0}}} sinn 
\biggl\{ \sqrt{\Omega_{k0}} \int_0^z dz'[\Omega_\gamma (z')
\nonumber \\
&& 
+ (\Omega_{\scriptscriptstyle DM}(z')
 + \Omega_B(z')) 
\nonumber \\
&& 
 + \Omega_{k0}(1 + z')^2
 + \Omega_\Lambda+  \Omega_{\chi}(z') \biggr]^{-1/2}  \biggr\}~~,
\end{eqnarray}
where $H_0$ is the present value of the Hubble constant, and $sinn(x) =
\sinh{x}$ for $\Omega_{k0} > 0$, $sinn(x)=x$, for  $\Omega_{k0} = 0$  and
$sinn(x) = \sin{x}$ for $\Omega_{k0} < 0$. 
The $\Omega_i$ are the energy densities normalized by the current
critical density, i.e. $\Omega_i(z) = {8 \pi G \rho_i(z) / 3 H_0^2}$.
Note the $\Omega_{DM}$ has a nontrivial redshift dependence in
the present cosmology via Eq.~(\ref{rho-DM}).

The look-back time $t_0 - t$ is a function of redshift becomes,
\begin{eqnarray}
t_0 -  t &=&  H_0^{-1}  \biggr\{ \int_0^z (1 + z')^{-1} 
 \biggl[\Omega_\gamma (z')
 \nonumber \\
& +& \Omega_{B}(z') +  \Omega_{\scriptscriptstyle DM} (z') 
\nonumber \\
 &+& \Omega_{k0}(1 + z')^2
 + \Omega_\Lambda 
 + \Omega_{\chi}(z')
\biggr]^{-1/2} dz' \biggr\}~~.
\end{eqnarray}

We have found  the best fits to the supernova magnitude-redshift
relation
by maximizing the  likelihood functions in  an 
effective $\chi^2$ analysis,
\begin{eqnarray}
 -2\ln {\cal L}_{SN} &=& \chi^2_{\rm eff} \nonumber \\
 &=& \sum{\frac{(Y_i^{data} - Y_i^{calc})^2}{\sigma_i^2}} \nonumber \\
 &-& \left(\sum \frac{Y_i^{data} - Y_i^{calc}}{\sigma_i^2}\right)^2/\sum\frac{1}{\sigma_i^2}.
\end{eqnarray}
where the second term corresponds to and
 analytic marginalization over the
absolute magnitude of the SNIa data with a flat prior \cite{Bridle:2001zv}.

\begin{figure}[t]
\begin{center}
\rotatebox{-90}{\includegraphics[width=6.5cm]{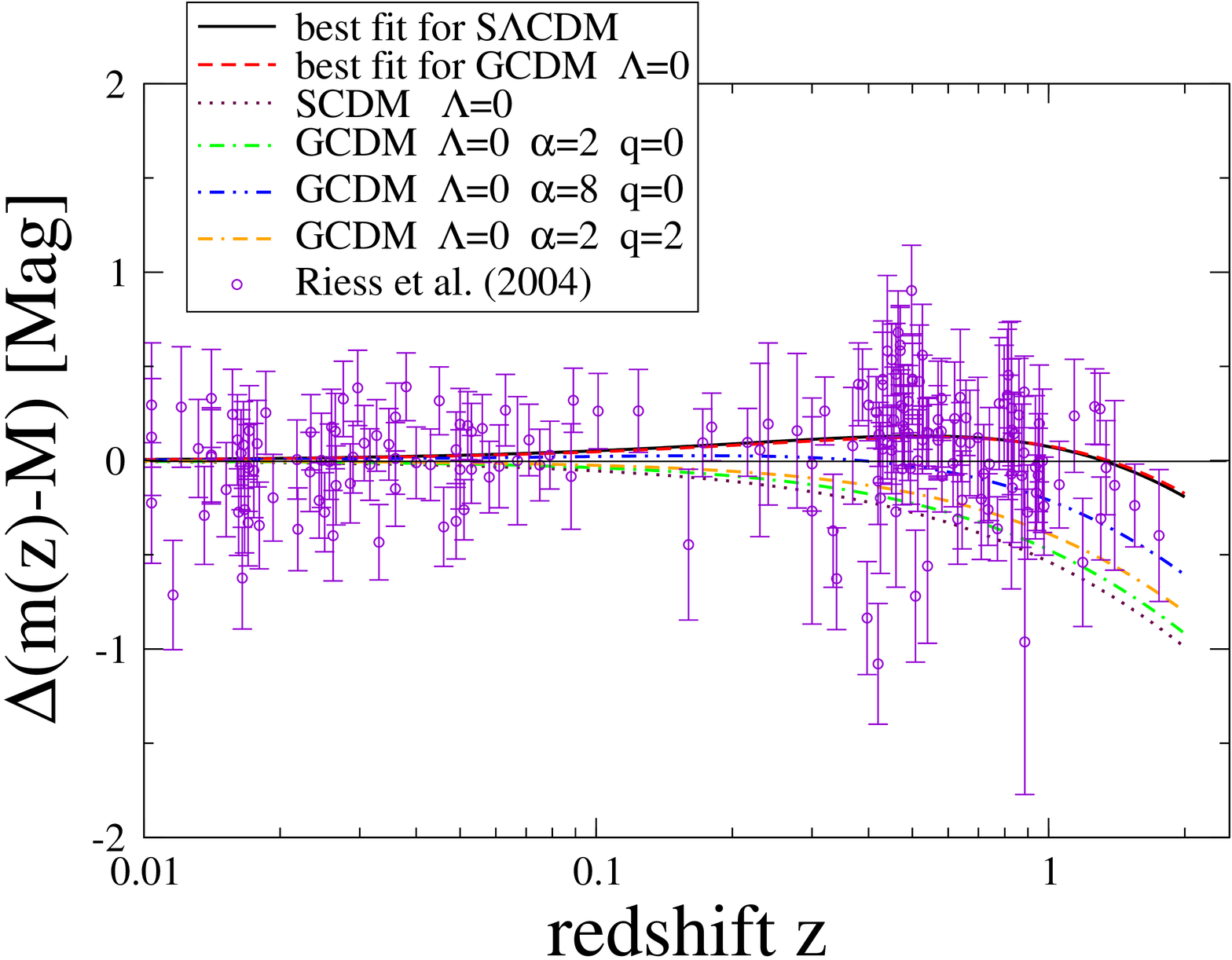}}
\end{center}
\begin{center}
\rotatebox{-90}{\includegraphics[width=6.5cm]{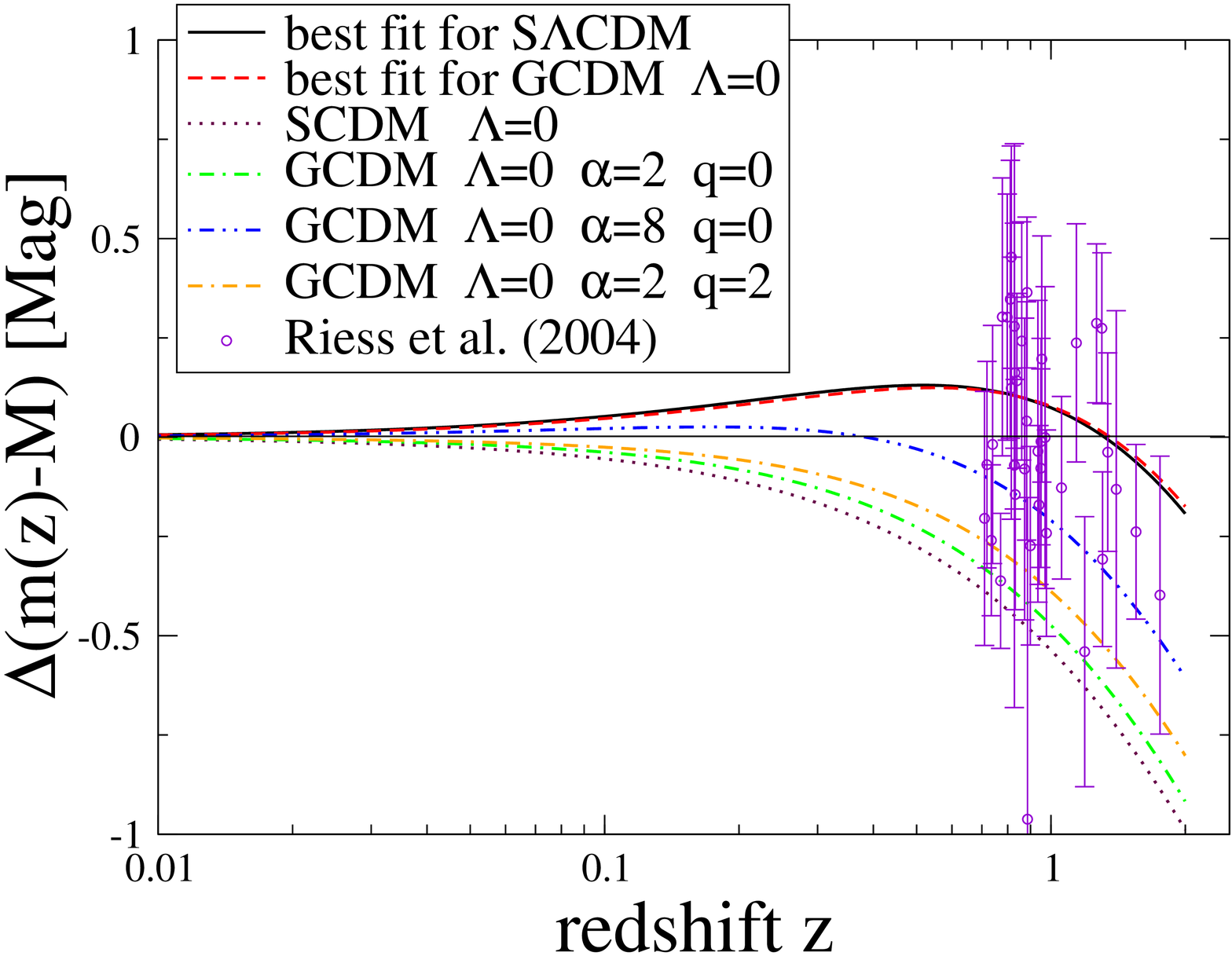}}
\end{center}
\vspace{-0.5cm}
\caption{Illustration of the supernovae magnitude redshift relation with and
 without GCDM.  The upper figure shows all data set of \cite{Riess:1998cb}. The
 lower figure highlights the points with $z > 0.7$ most relevant
 to this paper. Note, that the best fit model is nearly indistinguishable
 from the  S$\Lambda$CDM model.}
\label{fig:mag}
\end{figure}

Figure \ref{fig:mag} compares various cosmological models with some of the
recent combined data from the High-Z Supernova Search Team and the Supernova
Cosmology Project \cite{garnavich,Riess:1998cb,Perlmutter:1998np}. The lower
figure highlights the crucial data points at the highest redshift which
constrain the redshift evolution during the dark-matter dominated decelerating
phase. Shown are the K-corrected magnitudes  $m = M + 5 \log{ D_L} + 25$
vs.~redshift.

Curves are plotted relative to an open $\Omega_{\scriptscriptstyle DM},
\Omega_B , \Omega_{\Lambda}, \Omega_{\chi} = 0$, $\Omega_k  = 1$ cosmology. 

The best-fit growing cold dark matter (GCDM) models shown on this figure
correspond to either the constant $U_5$ or $U_5 = l H$ models.
They are indistinguishable from each other on this plot.  Of particular interest on Figure \ref{fig:mag} is the fact that our 
best fit $\Lambda = 0$ growing cold dark matter (GCDM)
models are nearly indistinguishable from the best fit
 Standard $\Lambda + $cold dark matter (S$\Lambda$CDM)
 model. Thus, our model realizes the present cosmic
acceleration without a cosmological constant. 

This figure also illustrates an important point regarding the
EOS parameter for matter in the bulk dimension.
An accelerating cosmology requires that $\rho_\chi + \bar \rho$
be nearly constant.  This means that no matter what the EOS
for matter when it is on the brane, A constant $\rho_\chi + \bar \rho$
requires that $q $ be small in Eqs. (\ref{rho1}) to (\ref{rhochidot})
for matter in the bulk.    
This is illustrated by
the $q = 2$ curve of Figure \ref{fig:mag} which is unable to
reproduce the required cosmic acceleration.  For the SNIa
constraint alone, the best fit curve from Table 1 is for $q =0.006$.

\section{CMB MCMC Analysis}
The high-resolution measurement of the spectrum of 
temperature fluctuations in the cosmic microwave
background (CMB) by the Wilkinson Microwave Anisotropy Probe (WMAP) has become
one of the most stringent tests for cosmology. 
The standard procedure for obtaining cosmological parameters
\cite{Verde} is
based upon Bayesian statistics.  This approach gives  the
posterior probability distribution from which the optimal set
of the cosmological parameters and their confidence levels can
be deduced. 
In our analysis we followed the Markov Chain Monte Carlo (MCMC)
approach \cite{MCMC} and explored the likelihood in an eight dimensional
parameter space consisting of
six WMAP standard parameters ($\Omega_b h^2$, $\Omega_c
h^2$, $h$, $z_{\rm re}$, $n_s$, $A_s$) plus the  two brane-world
 parameters, $\alpha$ and $q$.

\begin{figure}[t]
\begin{center}
\rotatebox{-90}{\includegraphics[width=6.5cm]{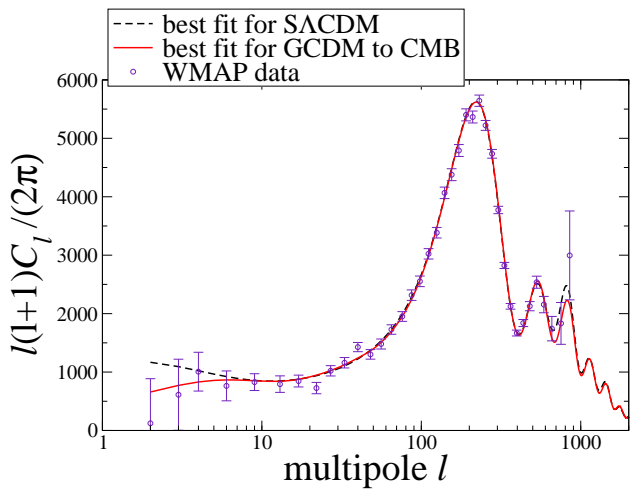}}
\end{center}
\vspace{-0.5cm}
 \caption{ CMB angular power spectrum with and without GCDM compared with
 observational data from WMAP. The dashed  line corresponds to the best
 fit  S$\Lambda$CDM model. The solid line shows 
 GCDM best-fit models with the 
 cosmological constant equal to zero, and $q=2.92$, $\alpha=2.14$. 
 This figure  demonstrates how GCDM modifies the CMB power spectrum.}
 \label{fig:Cls}
\end{figure}

Figure \ref{fig:Cls} shows best-fit  theoretical CMB 
power spectra along
with the combined data
set used to constrain our models. 
This figure
 illustrates how the 
flow of mass-energy exchange can modify the spectrum. 

There are essentially  two ways in which growing cold dark matter
models alter the CMB power spectrum. 
First, GCDM means that there is less dark matter at earlier
times. This leads to a smaller amplitude of the third acoustic
peak \cite{hu}. Second, the decay of the gravitational potential at late times 
is diminished due to the inflow of dark matter. This leads to a
smaller late integrated Sachs-Wolfe effect 
(LISW) and correspondingly
less power for the smallest multipoles.  

The integrated Sachs-Wolf effect accounts for changes
on the CMB anisotropy due  the time evolution of
the gravitational potentials as the photons  travel from
the last scattering surface to us.  
Writing the comoving metric perturbations as
\begin{equation}
ds^2 = a(\tau)^2[-(1+2 \Psi) d \tau^2 + (1+ 2 \Phi)\gamma_{i j} dx^i dx^j]
\end{equation}
For a photon in the absence of anisotropic stress 
$\Psi = \Phi$ and the ISW effect on the temperature power spectrum depends upon a simple  integral over $\dot \Phi $
 \cite{Liddle}.  Include anisotropic stress  we have:
\begin{equation}
\biggl(\frac{\delta T}{T}\biggr)_{ISW} = \int_{\tau_{lss}}^{\tau_0}
\biggl[  \dot \Psi - \dot \Phi \biggr] d\tau  ~~,
\end{equation}
where $d\tau = dt/a$ along the photon trajectory, and 
$ \dot \Phi =\partial \Phi/\partial \tau$.  

\begin{figure}[t]
\begin{center}
 \rotatebox{-90}{\includegraphics[width=6.5cm]{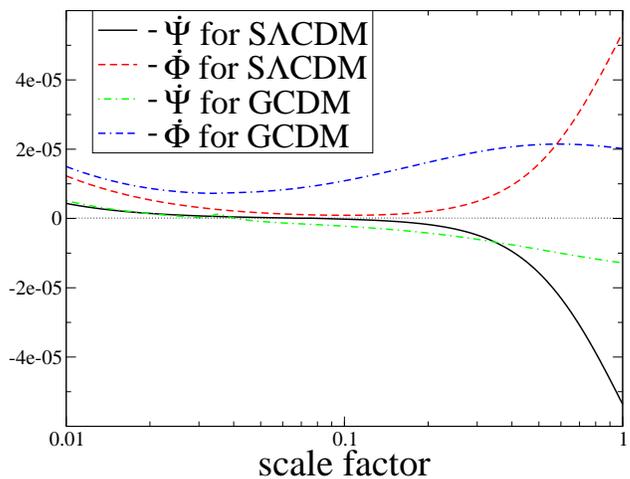}}
\end{center}
\vspace{-0.5cm}
\caption{Metric perturbations $ \dot \Psi$ and $\dot \Phi$ for the
integrated Sachs-Wolfe effect for the S$\Lambda$CDM and
GCDM models.}
\label{swolf}
\end{figure}

Figure \ref{swolf} illustrates
the evolution of the $\dot \Psi$ and $\dot \Phi$
potential derivatives with scale factor.  
In the $S\Lambda$CDM model the
gravitational potentials are rapidly decaying  at late times which causes
enhanced power on the largest scales (smallest multipoles).  In the GCDM model,
however, the potentials on the largest  scale  are changing less.
This accounts for the relative suppression of CMB power for the lowest multipoles.

For the combined SNIa and  CMB
data we again used
the MCMC approach in a seven dimensional
parameter space ($\Omega_b h^2$, $h_0$, $z_e$, $n_s$, $A_s$, $\alpha$, $q$).
It should be noted that in the GCDM cosmology of interest here
we must treat $\Omega_b$ and $\Omega_{DM}$ as
independent parameters in the SNIa analysis contrary to the usual
cosmologies.  This is because the evolution of the
energy densities $\rho_{b}$ and
$\rho_{DM}$ obey different functions of the cosmic scale factor $a$ in the
GCDM cosmology.

These data imply a minimum in $\chi^2$ of  1659.8 
for the GCDM model
without a cosmological constant, compared to a value of 1661.7  obtained
in the S$\Lambda$CDM
model. 
Note, however, that in our model there are seven parameters 
($\Omega_b h^2$,
$h_0$, $z_e$, $n_s$, $A_s$, $\alpha$, $q$), while 
in the S$\Lambda$CDM
model there are only six ($\Omega_b h^2$, $h_0$, $z_e$,
$n_s$, $A_s$, $\Omega_\Lambda$ ).  Hence,
there is no significant improvement in the reduced $\chi^2$ per degree
of freedom as evidenced in Table 1.

Although the GCDM  model explains the apparent
suppression of the CMB at low multipoles, the optical depth is rather large for the optimum fit to the CMB alone ($\tau=0.533$).  This is
due to a degeneracy among parameters so that a slightly better fit is obtained for the combination of a large $\tau$  offset by a smaller $h$ and $\Omega_{DM}$.   In
the combined fit with the SNIa data, however,  $h$ is better constrained so that a smaller  value of
$\tau=0.133$ results.  Note also, that in all of these fits
a large value of $\Omega_{DR} \sim 2-3$ is obtained.  This value
is offset, however, by the negative dark radiation component
as discussed below.  The key constraint is that $\Omega_{DM}+\Omega_{DR} \approx \Omega_{DM}+\Omega_{\Lambda}$ as evidenced in the fifth column of Table 1.

\vskip .2 in
\subsection{Matter Power Spectrum}
It is straight forward to determine the galactic matter power spectrum $P(k)$ to
compare with that deduced from large-scale structure surveys
\cite{sdss, 2df}.  For the usual power-law spectrum of primordial fluctuations
\begin{equation}
\delta^2_H(k) \propto k^{n-1}~~,
\end{equation}
a transfer function, $T(k)$ \cite{efstathiou} is required 
to convert from the amplitude of the perturbation as wave number $k$ enters
the horizon, to the present-day power spectrum, $P(k)$.
Transfer functions are 
easily computed using the code CMBFAST \cite{cmbfast} for various
sets of cosmological parameters. 

 An adequate approximate expression for the structure power
spectrum is
\begin{equation}
\frac{k^3}{2\pi^2}P(k) = \left( \frac{k}{aH_0} \right)^4 T^2(k)
\delta^2_H(k) ~~.
\end{equation}
This expression is only valid in the linear regime,
which in comoving wave number is up to approximately $k ^<_\sim
0.2~h$ Mpc$^{-1}$ and therefore
adequate for our purposes.
However, we also correct for
the nonlinear evolution of the power spectrum \cite{Peacock:1996ci}.

Here it is worth noting the way in which the power spectrum is
computed in our GCDM model.  It is necessary to specify the distribution of the
dark matter as it enters from the bulk dimension.
It would be most general  to suppose that there are fluctuations
in the bulk mass-energy distribution just as there are in the
brane.  Material entering from the bulk would then
appear with such fluctuations.  However, we make the simplifying assumption that the dark matter and dark
 radiation enter with uniform distributions.  After that, they
 are allowed to evolve as non-relativistic and relativistic particles, respectively,
 along with the already evolved structures on the brane.
 
 Hence, we define the matter over density as
\begin{equation}
\delta = \frac{\delta \rho_{DM}^{eff} + \delta \rho_M}
{\rho_{DM}^{eff} +  \rho_M} 
\end{equation}
where $\rho_{DM}^{eff}$ is the effective dark matter density as would be deduced, for example,  by the gravitational potential of a galaxy cluster.  For the GCDM model the effective over density is
\begin{equation} 
\delta \rho_{DM}^{eff} \equiv \delta \rho_{DM} + \delta \rho_{DR}
\end{equation}
and the effective gravitation mass energy density is
\begin{equation} 
\rho_{DM}^{eff} \equiv \rho_{DM} + \rho_{DR}~~.
\end{equation}
As usual, a bias parameter is introduced to account for
the difference between the galaxy and matter power spectrum,
\begin{equation}
\delta_{galaxy} = b \delta ~~.
\end{equation}

\begin{figure}[t]
\begin{center}
 \rotatebox{-90}{\includegraphics[width=6.5cm]{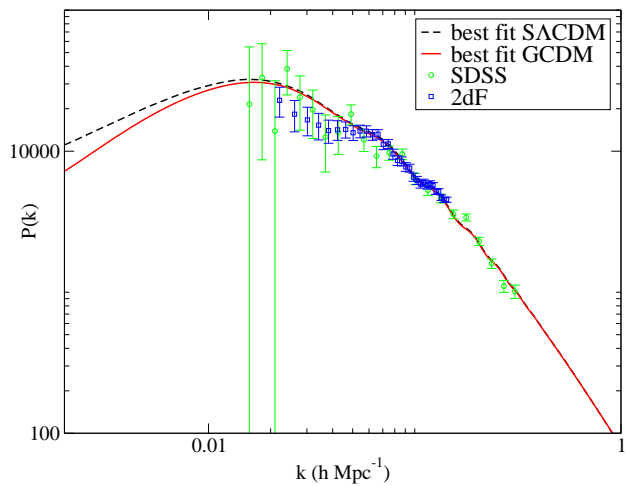}}
\end{center}
\vspace{-0.5cm}
\caption{Comparison of the matter power spectrum calculated
in the  $S\Lambda CDM$ cosmology with the
$GCDM$ cosmology which best fits the CMB + SNIa + P($k$) data.}
\label{fig:pk}\end{figure}

We have made a MCMC simultaneous fit to the CMB+SNIa+$P(k)$ data. We get the
best fit parameters, $q=0.037$ and $\alpha=8.33$.
Figure \ref{fig:pk} shows a comparison of the observed power spectrum with that of a  S$\Lambda$CDM model
and also the GCDM model which best fits the CMB+SNIa+$P(k)$ data.
The power spectrum derived in the best fit growing dark matter model
is almost indistinguishable from a  S$\Lambda$CDM model
until one gets to the very largest structures for which there is no data.
The parameters associated with this model are summarized in Table 1.  In these fits the bias $b$ is a marginalized parameter.
It is perhaps of note that the bias parameter deduced in this
way is somewhat larger $b = 2.1$ than that deduced in the
usual  S$\Lambda$CDM models, $b = 1.05$.
This derives from the fact that the dark matter potentials 
are not as deep at early times, and hence, galaxy formation
must be more efficient to produce the observed amplitude.

  \vskip .2 in

\ 
\section{Conclusion}
We have considered models in which the apparent cosmological constant derives from the 
energy exchange of cold dark matter from the bulk dimension 
to the brane.  The energy
momentum tensor in this extra dimensional brane cosmology
leads naturally to terms resembling a cosmological constant at the present time.  
If such energy exchange occurs our universe accelerates without the need to invoke a cosmological constant on the 
3-brane.

We find that such GCDM exchanges are consistent with 
observations including the supernova magnitude-redshift relation, temperature fluctuations in the CMB, and the matter power-spectrum data.    This cosmology is even slightly preferred as it fits better
the suppression of the CMB power spectrum at low multipoles.  We have thus demonstrated 
 that this cosmology represents an alternative model to the S$\Lambda$CDM cosmology for an observer on the 3-brane.  
A consistent fit to the observational constraints, however,
requires that  the EOS parameter for matter
in the bulk be small $q \approx 0.0$.  This EOS 
 is consistent with a need for
an AdS$_5$ geometry in the bulk.

 A peculiar feature of the present best fit models, 
is the fact that  the true value of $\Omega_{DM}$ is much larger 
than in the standard cosmology, though its gravitational effect is canceled by the dark-radiation contribution.  
Indeed, the key constraint is that $\Omega_{DM}+\Omega_{DR} \approx \Omega_{DM}+\Omega_{\Lambda}$ as evidenced in the fifth column of Table 1.

One consequence of such a large
and growing dark-matter contribution is the need for a somewhat large bias parameter.    
On the other hand, such a large dark matter content suggests new
observational tests of this cosmology.  For example,
if the dark-matter content
at the present time is much larger than in a  
S$\Lambda$CDM model, then direct terrestrial  measurements of the total density of cold dark-matter particles could indicate a much higher density than expected based upon their
mass and gravitation effect.  Another test of this cosmology is
that there should be a suppression of the matter power spectrum on the scale of the horizon compared to a  S$\Lambda$CDM cosmology.
Other tests of this paradigm are that a suppression of the third
acoustic peak in the CMB power spectrum occurs so that better data
near $l = 1000$ could help distinguish this cosmology.  We also
note that this cosmology produces large oscillations in the
CMB polarization power spectrum so that when better polarization
data are available it should help to eliminate of confirm this cosmology.
 
  An amusing feature of this model stems from the requirement
   that the present dark radiation from  in-flowing matter must cancel the effects of excess existing dark matter.  Hence, if the flow were to cease, the universe would change to a matter-dominated $\Omega_M 
   \approx 3$ cosmology which would begin  to collapse in about a
   hubble time.  In that sense, this is
  a new kind of doomsday cosmology \cite{doomsday}.

\acknowledgments

This work has been supported in part by the Mitsubishi Foundation,
the Grants-in-Aid for Scientific Research (13640313, 14540271) and for Specially Promoted Research (13002001) of the Ministry of Education, Science, Sports and Culture of Japan.  
K.I.'s work has been supported by a Grant-in-Aid for JSPS.
Work at the University of Notre Dame supported
by the U.S. Department of Energy under 
Nuclear Theory Grant DE-FG02-95-ER40934.

\centerline{ }
\centerline{ }
\centerline{ }
\centerline{ }
\centerline{ }
\centerline{ }
\vfill\eject
\centerline{ }
\centerline{ }
\centerline{ }
\centerline{ }

\newpage
\newpage
\vfill\eject
\pagebreak

~
~
~
~
~
~
~
~
~
~
~
~

~
\setlength{\oddsidemargin}{0cm}
\setlength{\voffset}{0pt}
\setlength{\topmargin}{0pt}
\setlength{\headheight}{0pt}
\setlength{\headsep}{5pt}
\setlength{\marginparwidth}{-3cm}
\setlength{\textheight}{23cm}
\setlength{\textwidth}{15.5cm}
\setlength{\footskip}{2cm}
\setlength{\marginparsep}{0pt}
\pagestyle{plain}

\begin{table}
\begin{center}
\begin{tabularx}{155mm}{lccccccccccccc}
\hline
 Fit  GCDM                  & $\alpha$ & q     & $\Omega_b h^2$ & $\Omega_{DM}+\Omega_{DR}$ & $\Omega_{DM}$ & $\Omega_{DR}$ & h    & $z_{re}$ & $n_s$ & $\tau$ & $b$ &$\chi^2_r$  \\ 
 Fit  S$\Lambda$CDM   &&&&$\Omega_{DM}+\Omega_{\Lambda}$ &  & $\Omega_{\Lambda}$ \\
 \hline
 SNIa Only              &          &       &                &                                           &                    &      &          &       &        &           \\ \hline
 Best Fit GCDM          & 11.0     & 0.006 & 0.022          &               0.93         & 3.31         & -2.38                & 0.58 & -        & -     & -      & - & 1.23     \\
 Best Fit S$\Lambda$CDM &    -     &     - & 0.022          &               0.97 & 0.26                        & 0.71               & 0.71 & -        & -     & -      & - & 1.24     \\ \hline

 CMB Only \\ \hline
 Best Fit GCDM          & 2.14     & 2.92  & 0.029          &               0.93         & 1.91         & -0.98                 & 0.64 & 29.1     & 1.18  & 0.533  & - & 1.02    \\
 Best Fit S$\Lambda$CDM &    -     &    -  & 0.023          &               0.94 & 0.23                        & 0.71               & 0.71 & 14.9     & 0.97  & 0.13   & - & 1.01    \\ \hline
 
 SNIa + CMB \\ \hline
 Best Fit GCDM          & 8.45     & 0.023 & 0.023          &               0.95         & 3.14         & -2.19                & 0.71 & 15.0     & 0.97  & 0.133  & - & 1.04    \\
 Best Fit S$\Lambda$CDM & -        & -     & 0.023          &               0.96 & 0.25                        & 0.71               & 0.70 & 13.3     & 0.96  & 0.111  & - & 1.04    \\ \hline

 SNIa + CMB + P(k) \\ \hline
 Best Fit GCDM          & 8.33     & 0.037 & 0.024          &               0.95         & 3.05         & -2.44                & 0.71 & 15.3     & 0.98  & 0.140  & 2.1 & 1.03    \\
 Best Fit S$\Lambda$CDM &    -     &     - & 0.023          &               0.95 & 0.24                        & 0.71               & 0.70 & 13.7     & 0.97  & 0.117  & 1.05 & 1.03    \\ \hline

\end{tabularx}
\end{center}
\caption{Parameter sets for various fits.}
\label{table_params}
\end{table}
\end{document}